\documentclass[12pt]{article}

\usepackage{amsmath}
\usepackage{graphicx}
\usepackage{amssymb}

\font\germanfont=eufm10 scaled\magstep1

\def\thefootnote{\fnsymbol{footnote}}
\def\baselinestretch{1.2}

\newtheorem{lemma}{Lemma}
\newtheorem{theorem}{Theorem}

\newtheorem{rem}{Remark}
\newtheorem{defi}{Definition}

\newcommand{\ch}{{\rm ch}}
\newcommand{\sh}{{\rm sh}}
\newcommand{\rmd}{{\rm d}}
{\end{eqnarray}\setcounter{equation}{\value{enumi}}}
\newcommand{\binomial}[2]{\left( \!
\begin{array}{c}
#1 \\ #2
\end{array}
\!\right)}

\renewcommand{\eqref}[1]{$(\ref{#1})$}
\renewcommand{\thesubsubsection}%
{\arabic{section}.\arabic{subsection}.\arabic{subsubsection}}

\renewcommand{\thesubsection}{\arabic{section}.\arabic{subsection}}
\renewcommand{\thesection}{\arabic{section}}

\makeatletter
\renewcommand{\thelemma}{\arabic{section}.\arabic{lemma}}%
\@addtoreset{lemma}{section}%
\renewcommand{\thetheorem}{\arabic{section}.\arabic{theorem}}%
\@addtoreset{theorem}{section}%
\renewcommand{\thecor}{\arabic{section}.\arabic{cor}}%
\@addtoreset{cor}{section}%
\renewcommand{\theequation}{\thesection.\arabic{equation}}%
\@addtoreset{equation}{section}%
\renewcommand{\theprop}{\thesection.\arabic{prop}}%
\@addtoreset{prop}{section}%
\makeatother

\newfont{\bg}{cmr10 scaled\magstep5}
\newcommand{\bigzerou}{\smash{\lower1.7ex\hbox{\bg 0}}}
\newcommand{\bigzerol}{\smash{\hbox{\bg 0}}}
\title{Fractional Logistic Map}
\author{Atsushi Nagai}
\date{}
\begin{document}
\begin{center}
{\bf\Large Fractional Logistic Map}\\
{\large Atsushi Nagai}\footnote{e-mail:a-nagai@sigmath.es.osaka-u.ac.jp}\\
{\it Graduate School of Engineering Science, Osaka University}\\
{\it 1-3 Matikaneyama, Toyonaka 560-8531, Japan}
\end{center}
\begin{abstract}
A new type of an integrable mapping is presented. This map is equipped
with fractional difference and possesses an exact solution, which can
be regarded as a discrete analogue of the Mittag-Leffler function.
\end{abstract}
\section{Introduction}
Logistic map~\cite{Morishita,Hirota5}
\begin{align}\label{eq:logistic}
\frac{u_{n+1} - u_n}{\varepsilon} = a u_n(1-u_{n+1}) \quad (a>0)
\end{align}
is an integrable discretization of the well-known logistic equation.
\begin{align}
\frac{d}{dt}u=a u(1-u) \quad (a>0)
\end{align}
Its origin was a population model in ecology. In addition, it has been
reported~\cite{Sato} that the logistic map is used in many other fields,
for example, agriculture, life sciences and engineering.

Through dependent variable transformation,
\begin{align}
u_n = \frac{1+a\varepsilon}
{\displaystyle a\varepsilon g_n+1+a\varepsilon}
\end{align}
the logistic map~\eqref{eq:logistic} is linearized as 
\begin{align}
& \Delta_{-n} g_{n}
= -a g_{n} \label{eq:linear1},
\end{align}
where $\Delta_{-n}$ is a backward difference operator defined by
\begin{align}
\Delta_{-n}=\varepsilon^{-1}(1-E^{-1}), E^{-1}f_n = f_{n-1}.
\end{align}
Since equation~\eqref{eq:linear1} possesses a solution,
\begin{align}
g_n = g_0(1+a\varepsilon)^{-n}
\end{align}
a solution to the logistic map~\eqref{eq:logistic} is given by
\begin{align}
u_n=\frac{u_0}
{u_0+(1-u_0)(1+a\varepsilon)^{-n}}
\end{align}

The main purpose of this paper is to provide a new type of an integrable
mapping, which can be regarded as an extension of the logistic map and 
is equipped with fractional difference. This map, which we call the 
fractional logistic map here, possesses an exact solution and have 
another parameter $p$ which corresponds to an order of difference.
In section 2, we introduce one definition of fractional difference,
which is a slight modification of Hirota's fractional difference
operator. We also find an eigenfunction of this operator. 
Section 3 is the main consequence of this paper and presents the 
fractional logistic map. We also show its time evolution through 
numerical experiment.

\section{Fractional difference}
We here give a definition of fractional difference operator and 
its eigenfunction. Before going to its definition,
let us introduce fundamental functions $M(a;n)$ defined by
\begin{align}\label{eq:dK0}
M(a;n)=\frac{1}{\Gamma(a)}\varepsilon^{a-1}
\frac{\Gamma(n+a-1)}{\Gamma(n)} 
=\varepsilon^{a-1}\binomial{n+a-2}{n-1}\quad (a>0, n\in{\Bbb Z}),
\end{align}
where $\varepsilon$ is an interval length and 
$\binomial{a}{n} \ (a\in{\Bbb R}, n\in{\Bbb Z})$ stands for
a binomial coefficient defined by
\begin{align*}
\binomial{a}{n}=
\begin{cases}
\displaystyle\frac{a(a-1)\cdots(a-n+1)}{n!}=
\displaystyle\frac{\Gamma(a+1)}{\Gamma(a-n+1)\Gamma(n+1)}
& (n>0)\\
1 & (n=0)\\
0 & (n<0)
\end{cases}
\end{align*}
This function satisfies the following lemma.
\begin{lemma}
The following relations hold.
\begin{align}
&M(a;0)=0 \quad (a>1)\label{eq:dK:-n}\\
&\Delta_{-n}{M}(a+1;n)=
\varepsilon^{-1}({M}(a+1;n)-{M}(a+1;n-1))=
{M}(a;n) \quad (a>0)\label{eq:dK1}
\end{align}
\end{lemma}
{\bf Proof}: Equation~\eqref{eq:dK:-n} is obvious owing to the definition
of binomial coefficient. Equation~\eqref{eq:dK1} is proved by using 
the relation $\Gamma(x+1)=x\Gamma(x)$ as follows.
\begin{align*}
\varepsilon^{-1}&(M(a+1;n)-M(a+1;n-1))
\\
&= \frac{\varepsilon^{a-1}}{\Gamma(a+1)}
\left(\frac{(n+a-1)\Gamma(n+a-1)}{\Gamma(n)}-
\frac{(n-1)\Gamma(n+a-1)}{\Gamma(n)}\right)\\
&= \frac{a \varepsilon^{a-1}}{\Gamma(a+1)}
\frac{\Gamma(n+a-1)}{\Gamma(n)}=M(a;n).
\end{align*}
\hfill$\blacksquare$

Next we go to the definition of fractional difference.
Hirota~\cite{Hirota} took the first $n$ terms of Taylor series of
$\Delta_{-n}^{\alpha}=\varepsilon^{-\alpha}(1-E^{-1})^{\alpha}$ and gave
the following definition.
\begin{defi}\label{def:dfrac1}
Let $\alpha \in {\Bbb R}$. Then
difference operator of order $\alpha$ is defined by
\begin{align}\label{eq:dfrac1}
\Delta_{-n}^\alpha u_n
=
\begin{cases}&\varepsilon^{-\alpha}
\displaystyle\sum_{j=0}^{n-1} \binomial{\alpha}{j}(-1)^j u_{n-j}\quad
\alpha\ne 1,2,\cdots\\
& \varepsilon^{-m}
\displaystyle\sum_{j=0}^{m} \binomial{m}{j}(-1)^j u_{n-j}\quad
\alpha=m\in{\Bbb Z}_{>0}
\end{cases}
\end{align}
\end{defi}
It should be noted that Diaz, Osler~\cite{Diaz} gave another definition
of fractional difference,
\begin{align}
\Delta_{-n}^\alpha u(t)&=\varepsilon^{-\alpha}
\sum_{j=0}^{\infty} \binomial{\alpha}{j}(-1)^j u_{n-j}
\end{align}
We here adopt another difference oparator $\Delta_{*,-n}^{\alpha}$ 
by modifying Hirota's operator.
\begin{defi}\label{def:dfrac2}
Let $\alpha\in{\Bbb R}$ and $m$ be an integer such that $m-1<\alpha\leq m$.
We define difference operator of order $\alpha$, $\Delta_{*,-n}^\alpha$,
by
\begin{align}
\label{eq:dfrac2}
 \Delta_{*,-n}^\alpha u_n \equiv \Delta_{-n}^{\alpha-m}
\Delta_{-n}^m u_n
= \varepsilon^{m-\alpha}\sum_{j=0}^{n-1} \binomial{\alpha-m}{j}
(-1)^j \Delta_{-k}^m u_k|_{k=n-j}
\end{align}
\end{defi}
We define a new function,
\begin{align}
F_a(\lambda,n)= \sum_{j=0}^\infty \lambda^jM(aj+1;n).
=\sum_{j=0}^\infty \lambda^j\varepsilon^{aj}
\frac{\Gamma(n+aj)}{\Gamma(aj+1)\Gamma(n)}
\end{align}
\begin{rem}
Putting $a=1$ in the above definition, we have
\begin{align}
F_1(\lambda,n)&=\sum_{j=0}^\infty \lambda^j\varepsilon^{j}
\frac{\Gamma(n+j)}{\Gamma(j+1)\Gamma(n)}\nonumber\\
&=\sum_{j=0}^\infty (\lambda\varepsilon)^{j}\binomial{n+j-1}{j}
\nonumber\\
&=\sum_{j=0}^\infty (-\lambda\varepsilon)^{j}\binomial{-n}{j}
=(1-\lambda\varepsilon)^{-n}.
\end{align}
\end{rem}
\begin{rem}
In the limit of $\varepsilon\rightarrow 0, n\rightarrow\infty$ with
$t=n\varepsilon$ fixed, the function $F_a(\lambda,n)$ converges to
\begin{align}
F_a(\lambda,n)\rightarrow \sum_{j=0}^\infty \frac{\lambda^j t^{aj}}
{\Gamma(aj+1)} = E_a(\lambda t^a).
\end{align}
$E_a(x)$ is the well-known Mittag-Leffler function and
and its asymptotic behavior is studied in detail in \cite{Sansonne}.
\end{rem}
\begin{theorem}\label{th:dML}
If $a>0$, the function $F_a(\lambda,n)$ is an eigen-function of the 
fractional difference operator \eqref{eq:dfrac2}. That is,
\begin{align}
\Delta_{*,-n}^a F_a(\lambda,n) = 
\lambda F_a(\lambda,n)
\end{align}
\end{theorem}
{\bf Proof of Theorem~\ref{th:dML}}: Let $m$ be an integer such that
$m-1<a\leq m$. Then we have
\begin{align}
\Delta_{*,-n}^a &F_a(\lambda,n)
= \Delta_{*,-n}^a \left(1+
\sum_{j=1}^\infty \lambda^j M(aj+1;n)\right)\nonumber\\
&= \Delta_{*,-n}^a \sum_{j=1}^\infty \lambda^j M(aj+1;n)\nonumber\\
&=\Delta_{-n}^{a-m}\sum_{j=1}^\infty \lambda^j \Delta_{-n}^m 
M(aj+1;n)\nonumber\\
&=\sum_{j=1}^\infty \lambda^j \Delta_{-n}^{a-m} M(aj+1-m;n)
\label{eq:proof1}
\end{align}
Each summand in the above equation is given by
\begin{align}
\Delta_{-n}^{a-m}&M(aj+1-m;n)
=\sum_{k=0}^{n-1}\binomial{a-m}{k}(-1)^k M(aj+1-m;n-k)\nonumber\\
&=\sum_{k=0}^{n-1}\binomial{a-m}{k}(-1)^k 
\binomial{aj-m+n-k-1}{n-k-1}\nonumber\\
&=\sum_{k=0}^{n-1}\binomial{a-m}{k}(-1)^{n-1}
\binomial{-aj+m-1}{n-k-1}\nonumber\\
&=(-1)^{n-1}\binomial{a-aj-1}{n-1}=\binomial{aj-a-1}{n-1}=M(aj-a+1;n).
\label{eq:proof2}
\end{align}
Therefore, substitution of eq.~\eqref{eq:proof2} into 
eq.~\eqref{eq:proof1} gives
\begin{align}
\Delta_{*,-n}^a F_a(\lambda,n)
&=\sum_{j=1}^\infty \lambda^j M(aj-a+1;n)
=\sum_{j=0}^\infty \lambda^{j+1} M(aj+1;n)=\lambda F_a(\lambda,n)
\end{align}
which completes the proof.\hfill$\blacksquare$

\section{Fractional Logistic map}
This section presents a fractionl integrable mapping, which we call
fractional logistic map here. We start with a linear equation
\begin{align}\label{eq:linearp}
& \Delta_{*,-n}^p g_n = -a g_n \quad (0<p\leq 1, a>0)
\end{align}
instead of linear eq.~\eqref{eq:linear1}. The above equation is 
rewritten as
\begin{align}
g_n = (1+a\varepsilon^p)^{-1}\left\{g_{n-1}-
\sum_{j=1}^{n-1} \binomial{p-1}{j}(-1)^j (g_{n-j}-g_{n-j-1})\right\}
\end{align}
Through the similar dependent variable transformation as logistic map,
\begin{align}
u_n = \frac{1+a\varepsilon^p}{a\varepsilon^p g_n +1+a\varepsilon^p}
\end{align}
we finally obtain a fractional logistic map
\begin{align}
\label{eq:flm}
u_n = \frac{1}{1+\displaystyle\frac{1}{1+a\varepsilon^p}\left\{
\frac{1}{u_{n-1}}-
\sum_{j=1}^{n-1} \binomial{p-1}{j}(-1)^j \left(\frac{1}{u_{n-j}}
-\frac{1}{u_{n-j-1}}\right)\right\}}.
\end{align}
Due to the Theorem~\ref{th:dML}, eq.~\eqref{eq:linearp} has a solution,
\begin{align}
g_n = g_0 F_p(-a;n).
\end{align}
Therefore, a solution to the fractional logistic map~\eqref{eq:flm}
is given by
\begin{align}
u_n=\frac{u_0}
{u_0+(1-u_0)F_p(-a;n)}
\end{align}
Putting $p=1$ in eq.~\eqref{eq:flm}, we have
\begin{align}
u_n = \frac{(1+a\varepsilon)u_{n-1}}{1+(1+a\varepsilon)u_{n-1}},
\end{align}
which recovers the logistic map. The following figure illustrates
time evolutions of fractional logistic map with order parameter $p = n/4
(n=1,2,3,4)$. We have put $u_0=0.1$, $a=1.0$
and $\varepsilon=0.1$. 

\begin{figure}[htbp]
\begin{center}
\setlength{\unitlength}{0.240900pt}
\ifx\plotpoint\undefined\newsavebox{\plotpoint}\fi
\sbox{\plotpoint}{\rule[-0.200pt]{0.400pt}{0.400pt}}%
\begin{picture}(1500,900)(0,0)
\font\gnuplot=cmr10 at 10pt
\gnuplot
\sbox{\plotpoint}{\rule[-0.200pt]{0.400pt}{0.400pt}}%
\put(161.0,123.0){\rule[-0.200pt]{4.818pt}{0.400pt}}
\put(141,123){\makebox(0,0)[r]{0}}
\put(1419.0,123.0){\rule[-0.200pt]{4.818pt}{0.400pt}}
\put(161.0,221.0){\rule[-0.200pt]{4.818pt}{0.400pt}}
\put(141,221){\makebox(0,0)[r]{0.2}}
\put(1419.0,221.0){\rule[-0.200pt]{4.818pt}{0.400pt}}
\put(161.0,320.0){\rule[-0.200pt]{4.818pt}{0.400pt}}
\put(141,320){\makebox(0,0)[r]{0.4}}
\put(1419.0,320.0){\rule[-0.200pt]{4.818pt}{0.400pt}}
\put(161.0,418.0){\rule[-0.200pt]{4.818pt}{0.400pt}}
\put(141,418){\makebox(0,0)[r]{0.6}}
\put(1419.0,418.0){\rule[-0.200pt]{4.818pt}{0.400pt}}
\put(161.0,516.0){\rule[-0.200pt]{4.818pt}{0.400pt}}
\put(141,516){\makebox(0,0)[r]{0.8}}
\put(1419.0,516.0){\rule[-0.200pt]{4.818pt}{0.400pt}}
\put(161.0,614.0){\rule[-0.200pt]{4.818pt}{0.400pt}}
\put(141,614){\makebox(0,0)[r]{1}}
\put(1419.0,614.0){\rule[-0.200pt]{4.818pt}{0.400pt}}
\put(161.0,713.0){\rule[-0.200pt]{4.818pt}{0.400pt}}
\put(141,713){\makebox(0,0)[r]{1.2}}
\put(1419.0,713.0){\rule[-0.200pt]{4.818pt}{0.400pt}}
\put(161.0,811.0){\rule[-0.200pt]{4.818pt}{0.400pt}}
\put(141,811){\makebox(0,0)[r]{1.4}}
\put(1419.0,811.0){\rule[-0.200pt]{4.818pt}{0.400pt}}
\put(161.0,123.0){\rule[-0.200pt]{0.400pt}{4.818pt}}
\put(161,82){\makebox(0,0){0}}
\put(161.0,840.0){\rule[-0.200pt]{0.400pt}{4.818pt}}
\put(480.0,123.0){\rule[-0.200pt]{0.400pt}{4.818pt}}
\put(480,82){\makebox(0,0){5}}
\put(480.0,840.0){\rule[-0.200pt]{0.400pt}{4.818pt}}
\put(800.0,123.0){\rule[-0.200pt]{0.400pt}{4.818pt}}
\put(800,82){\makebox(0,0){10}}
\put(800.0,840.0){\rule[-0.200pt]{0.400pt}{4.818pt}}
\put(1119.0,123.0){\rule[-0.200pt]{0.400pt}{4.818pt}}
\put(1119,82){\makebox(0,0){15}}
\put(1119.0,840.0){\rule[-0.200pt]{0.400pt}{4.818pt}}
\put(1439.0,123.0){\rule[-0.200pt]{0.400pt}{4.818pt}}
\put(1439,82){\makebox(0,0){20}}
\put(1439.0,840.0){\rule[-0.200pt]{0.400pt}{4.818pt}}
\put(161.0,123.0){\rule[-0.200pt]{0.400pt}{177.543pt}}
\put(161.0,123.0){\rule[-0.200pt]{307.870pt}{0.400pt}}
\put(1439.0,123.0){\rule[-0.200pt]{0.400pt}{177.543pt}}
\put(161.0,860.0){\rule[-0.200pt]{307.870pt}{0.400pt}}
\put(40,491){\makebox(0,0){$v_n$}}
\put(800,21){\makebox(0,0){$n$}}
\put(161.0,123.0){\rule[-0.200pt]{0.400pt}{177.543pt}}
\put(1279,820){\makebox(0,0)[r]{$p=1/4$}}
\put(1299.0,820.0){\rule[-0.200pt]{24.090pt}{0.400pt}}
\put(161,172){\usebox{\plotpoint}}
\multiput(161.59,172.00)(0.482,2.118){9}{\rule{0.116pt}{1.700pt}}
\multiput(160.17,172.00)(6.000,20.472){2}{\rule{0.400pt}{0.850pt}}
\multiput(167.00,196.59)(0.581,0.482){9}{\rule{0.567pt}{0.116pt}}
\multiput(167.00,195.17)(5.824,6.000){2}{\rule{0.283pt}{0.400pt}}
\multiput(174.00,202.61)(1.132,0.447){3}{\rule{0.900pt}{0.108pt}}
\multiput(174.00,201.17)(4.132,3.000){2}{\rule{0.450pt}{0.400pt}}
\multiput(180.00,205.61)(1.355,0.447){3}{\rule{1.033pt}{0.108pt}}
\multiput(180.00,204.17)(4.855,3.000){2}{\rule{0.517pt}{0.400pt}}
\put(187,208.17){\rule{1.300pt}{0.400pt}}
\multiput(187.00,207.17)(3.302,2.000){2}{\rule{0.650pt}{0.400pt}}
\put(193,210.17){\rule{1.300pt}{0.400pt}}
\multiput(193.00,209.17)(3.302,2.000){2}{\rule{0.650pt}{0.400pt}}
\put(199,212.17){\rule{1.500pt}{0.400pt}}
\multiput(199.00,211.17)(3.887,2.000){2}{\rule{0.750pt}{0.400pt}}
\put(206,213.67){\rule{1.445pt}{0.400pt}}
\multiput(206.00,213.17)(3.000,1.000){2}{\rule{0.723pt}{0.400pt}}
\put(212,214.67){\rule{1.686pt}{0.400pt}}
\multiput(212.00,214.17)(3.500,1.000){2}{\rule{0.843pt}{0.400pt}}
\put(219,215.67){\rule{1.445pt}{0.400pt}}
\multiput(219.00,215.17)(3.000,1.000){2}{\rule{0.723pt}{0.400pt}}
\put(225,216.67){\rule{1.445pt}{0.400pt}}
\multiput(225.00,216.17)(3.000,1.000){2}{\rule{0.723pt}{0.400pt}}
\put(231,217.67){\rule{1.686pt}{0.400pt}}
\multiput(231.00,217.17)(3.500,1.000){2}{\rule{0.843pt}{0.400pt}}
\put(238,218.67){\rule{1.445pt}{0.400pt}}
\multiput(238.00,218.17)(3.000,1.000){2}{\rule{0.723pt}{0.400pt}}
\put(244,219.67){\rule{1.445pt}{0.400pt}}
\multiput(244.00,219.17)(3.000,1.000){2}{\rule{0.723pt}{0.400pt}}
\put(250,220.67){\rule{1.686pt}{0.400pt}}
\multiput(250.00,220.17)(3.500,1.000){2}{\rule{0.843pt}{0.400pt}}
\put(257,221.67){\rule{1.445pt}{0.400pt}}
\multiput(257.00,221.17)(3.000,1.000){2}{\rule{0.723pt}{0.400pt}}
\put(270,222.67){\rule{1.445pt}{0.400pt}}
\multiput(270.00,222.17)(3.000,1.000){2}{\rule{0.723pt}{0.400pt}}
\put(276,223.67){\rule{1.445pt}{0.400pt}}
\multiput(276.00,223.17)(3.000,1.000){2}{\rule{0.723pt}{0.400pt}}
\put(263.0,223.0){\rule[-0.200pt]{1.686pt}{0.400pt}}
\put(289,224.67){\rule{1.445pt}{0.400pt}}
\multiput(289.00,224.17)(3.000,1.000){2}{\rule{0.723pt}{0.400pt}}
\put(295,225.67){\rule{1.686pt}{0.400pt}}
\multiput(295.00,225.17)(3.500,1.000){2}{\rule{0.843pt}{0.400pt}}
\put(282.0,225.0){\rule[-0.200pt]{1.686pt}{0.400pt}}
\put(308,226.67){\rule{1.445pt}{0.400pt}}
\multiput(308.00,226.17)(3.000,1.000){2}{\rule{0.723pt}{0.400pt}}
\put(302.0,227.0){\rule[-0.200pt]{1.445pt}{0.400pt}}
\put(321,227.67){\rule{1.445pt}{0.400pt}}
\multiput(321.00,227.17)(3.000,1.000){2}{\rule{0.723pt}{0.400pt}}
\put(314.0,228.0){\rule[-0.200pt]{1.686pt}{0.400pt}}
\put(334,228.67){\rule{1.445pt}{0.400pt}}
\multiput(334.00,228.17)(3.000,1.000){2}{\rule{0.723pt}{0.400pt}}
\put(327.0,229.0){\rule[-0.200pt]{1.686pt}{0.400pt}}
\put(346,229.67){\rule{1.686pt}{0.400pt}}
\multiput(346.00,229.17)(3.500,1.000){2}{\rule{0.843pt}{0.400pt}}
\put(340.0,230.0){\rule[-0.200pt]{1.445pt}{0.400pt}}
\put(359,230.67){\rule{1.445pt}{0.400pt}}
\multiput(359.00,230.17)(3.000,1.000){2}{\rule{0.723pt}{0.400pt}}
\put(353.0,231.0){\rule[-0.200pt]{1.445pt}{0.400pt}}
\put(378,231.67){\rule{1.686pt}{0.400pt}}
\multiput(378.00,231.17)(3.500,1.000){2}{\rule{0.843pt}{0.400pt}}
\put(365.0,232.0){\rule[-0.200pt]{3.132pt}{0.400pt}}
\put(397,232.67){\rule{1.686pt}{0.400pt}}
\multiput(397.00,232.17)(3.500,1.000){2}{\rule{0.843pt}{0.400pt}}
\put(385.0,233.0){\rule[-0.200pt]{2.891pt}{0.400pt}}
\put(410,233.67){\rule{1.686pt}{0.400pt}}
\multiput(410.00,233.17)(3.500,1.000){2}{\rule{0.843pt}{0.400pt}}
\put(404.0,234.0){\rule[-0.200pt]{1.445pt}{0.400pt}}
\put(429,234.67){\rule{1.686pt}{0.400pt}}
\multiput(429.00,234.17)(3.500,1.000){2}{\rule{0.843pt}{0.400pt}}
\put(417.0,235.0){\rule[-0.200pt]{2.891pt}{0.400pt}}
\put(449,235.67){\rule{1.445pt}{0.400pt}}
\multiput(449.00,235.17)(3.000,1.000){2}{\rule{0.723pt}{0.400pt}}
\put(436.0,236.0){\rule[-0.200pt]{3.132pt}{0.400pt}}
\put(474,236.67){\rule{1.445pt}{0.400pt}}
\multiput(474.00,236.17)(3.000,1.000){2}{\rule{0.723pt}{0.400pt}}
\put(455.0,237.0){\rule[-0.200pt]{4.577pt}{0.400pt}}
\put(493,237.67){\rule{1.686pt}{0.400pt}}
\multiput(493.00,237.17)(3.500,1.000){2}{\rule{0.843pt}{0.400pt}}
\put(480.0,238.0){\rule[-0.200pt]{3.132pt}{0.400pt}}
\put(519,238.67){\rule{1.445pt}{0.400pt}}
\multiput(519.00,238.17)(3.000,1.000){2}{\rule{0.723pt}{0.400pt}}
\put(500.0,239.0){\rule[-0.200pt]{4.577pt}{0.400pt}}
\put(544,239.67){\rule{1.686pt}{0.400pt}}
\multiput(544.00,239.17)(3.500,1.000){2}{\rule{0.843pt}{0.400pt}}
\put(525.0,240.0){\rule[-0.200pt]{4.577pt}{0.400pt}}
\put(570,240.67){\rule{1.445pt}{0.400pt}}
\multiput(570.00,240.17)(3.000,1.000){2}{\rule{0.723pt}{0.400pt}}
\put(551.0,241.0){\rule[-0.200pt]{4.577pt}{0.400pt}}
\put(596,241.67){\rule{1.445pt}{0.400pt}}
\multiput(596.00,241.17)(3.000,1.000){2}{\rule{0.723pt}{0.400pt}}
\put(576.0,242.0){\rule[-0.200pt]{4.818pt}{0.400pt}}
\put(627,242.67){\rule{1.686pt}{0.400pt}}
\multiput(627.00,242.17)(3.500,1.000){2}{\rule{0.843pt}{0.400pt}}
\put(602.0,243.0){\rule[-0.200pt]{6.022pt}{0.400pt}}
\put(659,243.67){\rule{1.686pt}{0.400pt}}
\multiput(659.00,243.17)(3.500,1.000){2}{\rule{0.843pt}{0.400pt}}
\put(634.0,244.0){\rule[-0.200pt]{6.022pt}{0.400pt}}
\put(691,244.67){\rule{1.686pt}{0.400pt}}
\multiput(691.00,244.17)(3.500,1.000){2}{\rule{0.843pt}{0.400pt}}
\put(666.0,245.0){\rule[-0.200pt]{6.022pt}{0.400pt}}
\put(723,245.67){\rule{1.686pt}{0.400pt}}
\multiput(723.00,245.17)(3.500,1.000){2}{\rule{0.843pt}{0.400pt}}
\put(698.0,246.0){\rule[-0.200pt]{6.022pt}{0.400pt}}
\put(762,246.67){\rule{1.445pt}{0.400pt}}
\multiput(762.00,246.17)(3.000,1.000){2}{\rule{0.723pt}{0.400pt}}
\put(730.0,247.0){\rule[-0.200pt]{7.709pt}{0.400pt}}
\put(800,247.67){\rule{1.445pt}{0.400pt}}
\multiput(800.00,247.17)(3.000,1.000){2}{\rule{0.723pt}{0.400pt}}
\put(768.0,248.0){\rule[-0.200pt]{7.709pt}{0.400pt}}
\put(838,248.67){\rule{1.686pt}{0.400pt}}
\multiput(838.00,248.17)(3.500,1.000){2}{\rule{0.843pt}{0.400pt}}
\put(806.0,249.0){\rule[-0.200pt]{7.709pt}{0.400pt}}
\put(883,249.67){\rule{1.445pt}{0.400pt}}
\multiput(883.00,249.17)(3.000,1.000){2}{\rule{0.723pt}{0.400pt}}
\put(845.0,250.0){\rule[-0.200pt]{9.154pt}{0.400pt}}
\put(928,250.67){\rule{1.445pt}{0.400pt}}
\multiput(928.00,250.17)(3.000,1.000){2}{\rule{0.723pt}{0.400pt}}
\put(889.0,251.0){\rule[-0.200pt]{9.395pt}{0.400pt}}
\put(973,251.67){\rule{1.445pt}{0.400pt}}
\multiput(973.00,251.17)(3.000,1.000){2}{\rule{0.723pt}{0.400pt}}
\put(934.0,252.0){\rule[-0.200pt]{9.395pt}{0.400pt}}
\put(1024,252.67){\rule{1.445pt}{0.400pt}}
\multiput(1024.00,252.17)(3.000,1.000){2}{\rule{0.723pt}{0.400pt}}
\put(979.0,253.0){\rule[-0.200pt]{10.840pt}{0.400pt}}
\put(1075,253.67){\rule{1.445pt}{0.400pt}}
\multiput(1075.00,253.17)(3.000,1.000){2}{\rule{0.723pt}{0.400pt}}
\put(1030.0,254.0){\rule[-0.200pt]{10.840pt}{0.400pt}}
\put(1132,254.67){\rule{1.686pt}{0.400pt}}
\multiput(1132.00,254.17)(3.500,1.000){2}{\rule{0.843pt}{0.400pt}}
\put(1081.0,255.0){\rule[-0.200pt]{12.286pt}{0.400pt}}
\put(1190,255.67){\rule{1.445pt}{0.400pt}}
\multiput(1190.00,255.17)(3.000,1.000){2}{\rule{0.723pt}{0.400pt}}
\put(1139.0,256.0){\rule[-0.200pt]{12.286pt}{0.400pt}}
\put(1247,256.67){\rule{1.686pt}{0.400pt}}
\multiput(1247.00,256.17)(3.500,1.000){2}{\rule{0.843pt}{0.400pt}}
\put(1196.0,257.0){\rule[-0.200pt]{12.286pt}{0.400pt}}
\put(1311,257.67){\rule{1.686pt}{0.400pt}}
\multiput(1311.00,257.17)(3.500,1.000){2}{\rule{0.843pt}{0.400pt}}
\put(1254.0,258.0){\rule[-0.200pt]{13.731pt}{0.400pt}}
\put(1381,258.67){\rule{1.686pt}{0.400pt}}
\multiput(1381.00,258.17)(3.500,1.000){2}{\rule{0.843pt}{0.400pt}}
\put(1318.0,259.0){\rule[-0.200pt]{15.177pt}{0.400pt}}
\put(1388.0,260.0){\rule[-0.200pt]{12.286pt}{0.400pt}}
\put(1279,779){\makebox(0,0)[r]{$p=1/2$}}
\multiput(1299,779)(20.756,0.000){5}{\usebox{\plotpoint}}
\put(1399,779){\usebox{\plotpoint}}
\put(161,172){\usebox{\plotpoint}}
\put(161.00,172.00){\usebox{\plotpoint}}
\put(170.91,189.91){\usebox{\plotpoint}}
\put(186.85,202.92){\usebox{\plotpoint}}
\put(204.67,213.43){\usebox{\plotpoint}}
\put(223.46,222.23){\usebox{\plotpoint}}
\put(242.78,229.59){\usebox{\plotpoint}}
\put(262.25,236.75){\usebox{\plotpoint}}
\put(282.26,242.08){\usebox{\plotpoint}}
\put(302.34,247.11){\usebox{\plotpoint}}
\put(322.32,252.44){\usebox{\plotpoint}}
\put(342.53,256.84){\usebox{\plotpoint}}
\put(362.65,261.61){\usebox{\plotpoint}}
\put(383.02,265.43){\usebox{\plotpoint}}
\put(403.47,268.92){\usebox{\plotpoint}}
\put(423.97,272.16){\usebox{\plotpoint}}
\put(444.40,275.69){\usebox{\plotpoint}}
\put(464.85,278.55){\usebox{\plotpoint}}
\put(485.35,281.76){\usebox{\plotpoint}}
\put(505.86,284.98){\usebox{\plotpoint}}
\put(526.38,288.00){\usebox{\plotpoint}}
\put(546.94,290.42){\usebox{\plotpoint}}
\put(567.52,292.59){\usebox{\plotpoint}}
\put(588.02,295.84){\usebox{\plotpoint}}
\put(608.59,298.00){\usebox{\plotpoint}}
\put(629.16,300.31){\usebox{\plotpoint}}
\put(649.76,302.46){\usebox{\plotpoint}}
\put(670.34,304.72){\usebox{\plotpoint}}
\put(690.99,306.00){\usebox{\plotpoint}}
\put(711.58,308.10){\usebox{\plotpoint}}
\put(732.15,310.36){\usebox{\plotpoint}}
\put(752.73,312.62){\usebox{\plotpoint}}
\put(773.37,314.00){\usebox{\plotpoint}}
\put(793.99,316.00){\usebox{\plotpoint}}
\put(814.59,318.00){\usebox{\plotpoint}}
\put(835.23,319.54){\usebox{\plotpoint}}
\put(855.86,321.00){\usebox{\plotpoint}}
\put(876.47,322.92){\usebox{\plotpoint}}
\put(897.12,324.19){\usebox{\plotpoint}}
\put(917.73,326.00){\usebox{\plotpoint}}
\put(938.40,327.00){\usebox{\plotpoint}}
\put(959.01,328.86){\usebox{\plotpoint}}
\put(979.68,330.11){\usebox{\plotpoint}}
\put(1000.33,331.39){\usebox{\plotpoint}}
\put(1020.95,333.00){\usebox{\plotpoint}}
\put(1041.62,334.00){\usebox{\plotpoint}}
\put(1062.21,336.00){\usebox{\plotpoint}}
\put(1082.88,337.00){\usebox{\plotpoint}}
\put(1103.56,338.00){\usebox{\plotpoint}}
\put(1124.23,339.00){\usebox{\plotpoint}}
\put(1144.82,340.97){\usebox{\plotpoint}}
\put(1165.49,342.00){\usebox{\plotpoint}}
\put(1186.16,343.00){\usebox{\plotpoint}}
\put(1206.85,344.00){\usebox{\plotpoint}}
\put(1227.52,345.00){\usebox{\plotpoint}}
\put(1248.19,346.17){\usebox{\plotpoint}}
\put(1268.86,347.41){\usebox{\plotpoint}}
\put(1289.53,348.59){\usebox{\plotpoint}}
\put(1310.18,349.86){\usebox{\plotpoint}}
\put(1330.84,351.00){\usebox{\plotpoint}}
\put(1351.52,352.00){\usebox{\plotpoint}}
\put(1372.23,352.54){\usebox{\plotpoint}}
\put(1392.88,353.81){\usebox{\plotpoint}}
\put(1413.54,355.00){\usebox{\plotpoint}}
\put(1434.28,355.21){\usebox{\plotpoint}}
\put(1439,356){\usebox{\plotpoint}}
\sbox{\plotpoint}{\rule[-0.400pt]{0.800pt}{0.800pt}}%
\put(1279,738){\makebox(0,0)[r]{$p=3/4$}}
\put(1299.0,738.0){\rule[-0.400pt]{24.090pt}{0.800pt}}
\put(161,172){\usebox{\plotpoint}}
\multiput(162.39,172.00)(0.536,0.685){5}{\rule{0.129pt}{1.267pt}}
\multiput(159.34,172.00)(6.000,5.371){2}{\rule{0.800pt}{0.633pt}}
\multiput(167.00,181.39)(0.574,0.536){5}{\rule{1.133pt}{0.129pt}}
\multiput(167.00,178.34)(4.648,6.000){2}{\rule{0.567pt}{0.800pt}}
\multiput(174.00,187.39)(0.462,0.536){5}{\rule{1.000pt}{0.129pt}}
\multiput(174.00,184.34)(3.924,6.000){2}{\rule{0.500pt}{0.800pt}}
\multiput(180.00,193.39)(0.574,0.536){5}{\rule{1.133pt}{0.129pt}}
\multiput(180.00,190.34)(4.648,6.000){2}{\rule{0.567pt}{0.800pt}}
\multiput(187.00,199.38)(0.592,0.560){3}{\rule{1.160pt}{0.135pt}}
\multiput(187.00,196.34)(3.592,5.000){2}{\rule{0.580pt}{0.800pt}}
\multiput(193.00,204.38)(0.592,0.560){3}{\rule{1.160pt}{0.135pt}}
\multiput(193.00,201.34)(3.592,5.000){2}{\rule{0.580pt}{0.800pt}}
\multiput(199.00,209.39)(0.574,0.536){5}{\rule{1.133pt}{0.129pt}}
\multiput(199.00,206.34)(4.648,6.000){2}{\rule{0.567pt}{0.800pt}}
\multiput(206.00,215.38)(0.592,0.560){3}{\rule{1.160pt}{0.135pt}}
\multiput(206.00,212.34)(3.592,5.000){2}{\rule{0.580pt}{0.800pt}}
\multiput(212.00,220.38)(0.760,0.560){3}{\rule{1.320pt}{0.135pt}}
\multiput(212.00,217.34)(4.260,5.000){2}{\rule{0.660pt}{0.800pt}}
\multiput(219.00,225.38)(0.592,0.560){3}{\rule{1.160pt}{0.135pt}}
\multiput(219.00,222.34)(3.592,5.000){2}{\rule{0.580pt}{0.800pt}}
\multiput(225.00,230.38)(0.592,0.560){3}{\rule{1.160pt}{0.135pt}}
\multiput(225.00,227.34)(3.592,5.000){2}{\rule{0.580pt}{0.800pt}}
\put(231,234.34){\rule{1.600pt}{0.800pt}}
\multiput(231.00,232.34)(3.679,4.000){2}{\rule{0.800pt}{0.800pt}}
\multiput(238.00,239.38)(0.592,0.560){3}{\rule{1.160pt}{0.135pt}}
\multiput(238.00,236.34)(3.592,5.000){2}{\rule{0.580pt}{0.800pt}}
\multiput(244.00,244.38)(0.592,0.560){3}{\rule{1.160pt}{0.135pt}}
\multiput(244.00,241.34)(3.592,5.000){2}{\rule{0.580pt}{0.800pt}}
\put(250,248.34){\rule{1.600pt}{0.800pt}}
\multiput(250.00,246.34)(3.679,4.000){2}{\rule{0.800pt}{0.800pt}}
\multiput(257.00,253.38)(0.592,0.560){3}{\rule{1.160pt}{0.135pt}}
\multiput(257.00,250.34)(3.592,5.000){2}{\rule{0.580pt}{0.800pt}}
\put(263,257.34){\rule{1.600pt}{0.800pt}}
\multiput(263.00,255.34)(3.679,4.000){2}{\rule{0.800pt}{0.800pt}}
\put(270,261.34){\rule{1.400pt}{0.800pt}}
\multiput(270.00,259.34)(3.094,4.000){2}{\rule{0.700pt}{0.800pt}}
\multiput(276.00,266.38)(0.592,0.560){3}{\rule{1.160pt}{0.135pt}}
\multiput(276.00,263.34)(3.592,5.000){2}{\rule{0.580pt}{0.800pt}}
\put(282,270.34){\rule{1.600pt}{0.800pt}}
\multiput(282.00,268.34)(3.679,4.000){2}{\rule{0.800pt}{0.800pt}}
\put(289,274.34){\rule{1.400pt}{0.800pt}}
\multiput(289.00,272.34)(3.094,4.000){2}{\rule{0.700pt}{0.800pt}}
\put(295,278.34){\rule{1.600pt}{0.800pt}}
\multiput(295.00,276.34)(3.679,4.000){2}{\rule{0.800pt}{0.800pt}}
\put(302,282.34){\rule{1.400pt}{0.800pt}}
\multiput(302.00,280.34)(3.094,4.000){2}{\rule{0.700pt}{0.800pt}}
\put(308,286.34){\rule{1.400pt}{0.800pt}}
\multiput(308.00,284.34)(3.094,4.000){2}{\rule{0.700pt}{0.800pt}}
\put(314,290.34){\rule{1.600pt}{0.800pt}}
\multiput(314.00,288.34)(3.679,4.000){2}{\rule{0.800pt}{0.800pt}}
\put(321,293.84){\rule{1.445pt}{0.800pt}}
\multiput(321.00,292.34)(3.000,3.000){2}{\rule{0.723pt}{0.800pt}}
\put(327,297.34){\rule{1.600pt}{0.800pt}}
\multiput(327.00,295.34)(3.679,4.000){2}{\rule{0.800pt}{0.800pt}}
\put(334,301.34){\rule{1.400pt}{0.800pt}}
\multiput(334.00,299.34)(3.094,4.000){2}{\rule{0.700pt}{0.800pt}}
\put(340,304.84){\rule{1.445pt}{0.800pt}}
\multiput(340.00,303.34)(3.000,3.000){2}{\rule{0.723pt}{0.800pt}}
\put(346,308.34){\rule{1.600pt}{0.800pt}}
\multiput(346.00,306.34)(3.679,4.000){2}{\rule{0.800pt}{0.800pt}}
\put(353,311.84){\rule{1.445pt}{0.800pt}}
\multiput(353.00,310.34)(3.000,3.000){2}{\rule{0.723pt}{0.800pt}}
\put(359,314.84){\rule{1.445pt}{0.800pt}}
\multiput(359.00,313.34)(3.000,3.000){2}{\rule{0.723pt}{0.800pt}}
\put(365,317.84){\rule{1.686pt}{0.800pt}}
\multiput(365.00,316.34)(3.500,3.000){2}{\rule{0.843pt}{0.800pt}}
\put(372,321.34){\rule{1.400pt}{0.800pt}}
\multiput(372.00,319.34)(3.094,4.000){2}{\rule{0.700pt}{0.800pt}}
\put(378,324.84){\rule{1.686pt}{0.800pt}}
\multiput(378.00,323.34)(3.500,3.000){2}{\rule{0.843pt}{0.800pt}}
\put(385,327.84){\rule{1.445pt}{0.800pt}}
\multiput(385.00,326.34)(3.000,3.000){2}{\rule{0.723pt}{0.800pt}}
\put(391,330.84){\rule{1.445pt}{0.800pt}}
\multiput(391.00,329.34)(3.000,3.000){2}{\rule{0.723pt}{0.800pt}}
\put(397,333.84){\rule{1.686pt}{0.800pt}}
\multiput(397.00,332.34)(3.500,3.000){2}{\rule{0.843pt}{0.800pt}}
\put(404,336.34){\rule{1.445pt}{0.800pt}}
\multiput(404.00,335.34)(3.000,2.000){2}{\rule{0.723pt}{0.800pt}}
\put(410,338.84){\rule{1.686pt}{0.800pt}}
\multiput(410.00,337.34)(3.500,3.000){2}{\rule{0.843pt}{0.800pt}}
\put(417,341.84){\rule{1.445pt}{0.800pt}}
\multiput(417.00,340.34)(3.000,3.000){2}{\rule{0.723pt}{0.800pt}}
\put(423,344.84){\rule{1.445pt}{0.800pt}}
\multiput(423.00,343.34)(3.000,3.000){2}{\rule{0.723pt}{0.800pt}}
\put(429,347.34){\rule{1.686pt}{0.800pt}}
\multiput(429.00,346.34)(3.500,2.000){2}{\rule{0.843pt}{0.800pt}}
\put(436,349.84){\rule{1.445pt}{0.800pt}}
\multiput(436.00,348.34)(3.000,3.000){2}{\rule{0.723pt}{0.800pt}}
\put(442,352.34){\rule{1.686pt}{0.800pt}}
\multiput(442.00,351.34)(3.500,2.000){2}{\rule{0.843pt}{0.800pt}}
\put(449,354.84){\rule{1.445pt}{0.800pt}}
\multiput(449.00,353.34)(3.000,3.000){2}{\rule{0.723pt}{0.800pt}}
\put(455,357.34){\rule{1.445pt}{0.800pt}}
\multiput(455.00,356.34)(3.000,2.000){2}{\rule{0.723pt}{0.800pt}}
\put(461,359.84){\rule{1.686pt}{0.800pt}}
\multiput(461.00,358.34)(3.500,3.000){2}{\rule{0.843pt}{0.800pt}}
\put(468,362.34){\rule{1.445pt}{0.800pt}}
\multiput(468.00,361.34)(3.000,2.000){2}{\rule{0.723pt}{0.800pt}}
\put(474,364.34){\rule{1.445pt}{0.800pt}}
\multiput(474.00,363.34)(3.000,2.000){2}{\rule{0.723pt}{0.800pt}}
\put(480,366.84){\rule{1.686pt}{0.800pt}}
\multiput(480.00,365.34)(3.500,3.000){2}{\rule{0.843pt}{0.800pt}}
\put(487,369.34){\rule{1.445pt}{0.800pt}}
\multiput(487.00,368.34)(3.000,2.000){2}{\rule{0.723pt}{0.800pt}}
\put(493,371.34){\rule{1.686pt}{0.800pt}}
\multiput(493.00,370.34)(3.500,2.000){2}{\rule{0.843pt}{0.800pt}}
\put(500,373.34){\rule{1.445pt}{0.800pt}}
\multiput(500.00,372.34)(3.000,2.000){2}{\rule{0.723pt}{0.800pt}}
\put(506,375.34){\rule{1.445pt}{0.800pt}}
\multiput(506.00,374.34)(3.000,2.000){2}{\rule{0.723pt}{0.800pt}}
\put(512,377.34){\rule{1.686pt}{0.800pt}}
\multiput(512.00,376.34)(3.500,2.000){2}{\rule{0.843pt}{0.800pt}}
\put(519,379.34){\rule{1.445pt}{0.800pt}}
\multiput(519.00,378.34)(3.000,2.000){2}{\rule{0.723pt}{0.800pt}}
\put(525,381.34){\rule{1.686pt}{0.800pt}}
\multiput(525.00,380.34)(3.500,2.000){2}{\rule{0.843pt}{0.800pt}}
\put(532,383.34){\rule{1.445pt}{0.800pt}}
\multiput(532.00,382.34)(3.000,2.000){2}{\rule{0.723pt}{0.800pt}}
\put(538,385.34){\rule{1.445pt}{0.800pt}}
\multiput(538.00,384.34)(3.000,2.000){2}{\rule{0.723pt}{0.800pt}}
\put(544,387.34){\rule{1.686pt}{0.800pt}}
\multiput(544.00,386.34)(3.500,2.000){2}{\rule{0.843pt}{0.800pt}}
\put(551,389.34){\rule{1.445pt}{0.800pt}}
\multiput(551.00,388.34)(3.000,2.000){2}{\rule{0.723pt}{0.800pt}}
\put(557,390.84){\rule{1.686pt}{0.800pt}}
\multiput(557.00,390.34)(3.500,1.000){2}{\rule{0.843pt}{0.800pt}}
\put(564,392.34){\rule{1.445pt}{0.800pt}}
\multiput(564.00,391.34)(3.000,2.000){2}{\rule{0.723pt}{0.800pt}}
\put(570,394.34){\rule{1.445pt}{0.800pt}}
\multiput(570.00,393.34)(3.000,2.000){2}{\rule{0.723pt}{0.800pt}}
\put(576,396.34){\rule{1.686pt}{0.800pt}}
\multiput(576.00,395.34)(3.500,2.000){2}{\rule{0.843pt}{0.800pt}}
\put(583,397.84){\rule{1.445pt}{0.800pt}}
\multiput(583.00,397.34)(3.000,1.000){2}{\rule{0.723pt}{0.800pt}}
\put(589,399.34){\rule{1.686pt}{0.800pt}}
\multiput(589.00,398.34)(3.500,2.000){2}{\rule{0.843pt}{0.800pt}}
\put(596,401.34){\rule{1.445pt}{0.800pt}}
\multiput(596.00,400.34)(3.000,2.000){2}{\rule{0.723pt}{0.800pt}}
\put(602,402.84){\rule{1.445pt}{0.800pt}}
\multiput(602.00,402.34)(3.000,1.000){2}{\rule{0.723pt}{0.800pt}}
\put(608,404.34){\rule{1.686pt}{0.800pt}}
\multiput(608.00,403.34)(3.500,2.000){2}{\rule{0.843pt}{0.800pt}}
\put(615,405.84){\rule{1.445pt}{0.800pt}}
\multiput(615.00,405.34)(3.000,1.000){2}{\rule{0.723pt}{0.800pt}}
\put(621,407.34){\rule{1.445pt}{0.800pt}}
\multiput(621.00,406.34)(3.000,2.000){2}{\rule{0.723pt}{0.800pt}}
\put(627,408.84){\rule{1.686pt}{0.800pt}}
\multiput(627.00,408.34)(3.500,1.000){2}{\rule{0.843pt}{0.800pt}}
\put(634,410.34){\rule{1.445pt}{0.800pt}}
\multiput(634.00,409.34)(3.000,2.000){2}{\rule{0.723pt}{0.800pt}}
\put(640,411.84){\rule{1.686pt}{0.800pt}}
\multiput(640.00,411.34)(3.500,1.000){2}{\rule{0.843pt}{0.800pt}}
\put(647,413.34){\rule{1.445pt}{0.800pt}}
\multiput(647.00,412.34)(3.000,2.000){2}{\rule{0.723pt}{0.800pt}}
\put(653,414.84){\rule{1.445pt}{0.800pt}}
\multiput(653.00,414.34)(3.000,1.000){2}{\rule{0.723pt}{0.800pt}}
\put(659,415.84){\rule{1.686pt}{0.800pt}}
\multiput(659.00,415.34)(3.500,1.000){2}{\rule{0.843pt}{0.800pt}}
\put(666,417.34){\rule{1.445pt}{0.800pt}}
\multiput(666.00,416.34)(3.000,2.000){2}{\rule{0.723pt}{0.800pt}}
\put(672,418.84){\rule{1.686pt}{0.800pt}}
\multiput(672.00,418.34)(3.500,1.000){2}{\rule{0.843pt}{0.800pt}}
\put(679,419.84){\rule{1.445pt}{0.800pt}}
\multiput(679.00,419.34)(3.000,1.000){2}{\rule{0.723pt}{0.800pt}}
\put(685,421.34){\rule{1.445pt}{0.800pt}}
\multiput(685.00,420.34)(3.000,2.000){2}{\rule{0.723pt}{0.800pt}}
\put(691,422.84){\rule{1.686pt}{0.800pt}}
\multiput(691.00,422.34)(3.500,1.000){2}{\rule{0.843pt}{0.800pt}}
\put(698,423.84){\rule{1.445pt}{0.800pt}}
\multiput(698.00,423.34)(3.000,1.000){2}{\rule{0.723pt}{0.800pt}}
\put(704,424.84){\rule{1.686pt}{0.800pt}}
\multiput(704.00,424.34)(3.500,1.000){2}{\rule{0.843pt}{0.800pt}}
\put(711,426.34){\rule{1.445pt}{0.800pt}}
\multiput(711.00,425.34)(3.000,2.000){2}{\rule{0.723pt}{0.800pt}}
\put(717,427.84){\rule{1.445pt}{0.800pt}}
\multiput(717.00,427.34)(3.000,1.000){2}{\rule{0.723pt}{0.800pt}}
\put(723,428.84){\rule{1.686pt}{0.800pt}}
\multiput(723.00,428.34)(3.500,1.000){2}{\rule{0.843pt}{0.800pt}}
\put(730,429.84){\rule{1.445pt}{0.800pt}}
\multiput(730.00,429.34)(3.000,1.000){2}{\rule{0.723pt}{0.800pt}}
\put(736,430.84){\rule{1.445pt}{0.800pt}}
\multiput(736.00,430.34)(3.000,1.000){2}{\rule{0.723pt}{0.800pt}}
\put(742,431.84){\rule{1.686pt}{0.800pt}}
\multiput(742.00,431.34)(3.500,1.000){2}{\rule{0.843pt}{0.800pt}}
\put(749,432.84){\rule{1.445pt}{0.800pt}}
\multiput(749.00,432.34)(3.000,1.000){2}{\rule{0.723pt}{0.800pt}}
\put(755,434.34){\rule{1.686pt}{0.800pt}}
\multiput(755.00,433.34)(3.500,2.000){2}{\rule{0.843pt}{0.800pt}}
\put(762,435.84){\rule{1.445pt}{0.800pt}}
\multiput(762.00,435.34)(3.000,1.000){2}{\rule{0.723pt}{0.800pt}}
\put(768,436.84){\rule{1.445pt}{0.800pt}}
\multiput(768.00,436.34)(3.000,1.000){2}{\rule{0.723pt}{0.800pt}}
\put(774,437.84){\rule{1.686pt}{0.800pt}}
\multiput(774.00,437.34)(3.500,1.000){2}{\rule{0.843pt}{0.800pt}}
\put(781,438.84){\rule{1.445pt}{0.800pt}}
\multiput(781.00,438.34)(3.000,1.000){2}{\rule{0.723pt}{0.800pt}}
\put(787,439.84){\rule{1.686pt}{0.800pt}}
\multiput(787.00,439.34)(3.500,1.000){2}{\rule{0.843pt}{0.800pt}}
\put(794,440.84){\rule{1.445pt}{0.800pt}}
\multiput(794.00,440.34)(3.000,1.000){2}{\rule{0.723pt}{0.800pt}}
\put(800,441.84){\rule{1.445pt}{0.800pt}}
\multiput(800.00,441.34)(3.000,1.000){2}{\rule{0.723pt}{0.800pt}}
\put(806,442.84){\rule{1.686pt}{0.800pt}}
\multiput(806.00,442.34)(3.500,1.000){2}{\rule{0.843pt}{0.800pt}}
\put(813,443.84){\rule{1.445pt}{0.800pt}}
\multiput(813.00,443.34)(3.000,1.000){2}{\rule{0.723pt}{0.800pt}}
\put(819,444.84){\rule{1.686pt}{0.800pt}}
\multiput(819.00,444.34)(3.500,1.000){2}{\rule{0.843pt}{0.800pt}}
\put(826,445.84){\rule{1.445pt}{0.800pt}}
\multiput(826.00,445.34)(3.000,1.000){2}{\rule{0.723pt}{0.800pt}}
\put(832,446.84){\rule{1.445pt}{0.800pt}}
\multiput(832.00,446.34)(3.000,1.000){2}{\rule{0.723pt}{0.800pt}}
\put(845,447.84){\rule{1.445pt}{0.800pt}}
\multiput(845.00,447.34)(3.000,1.000){2}{\rule{0.723pt}{0.800pt}}
\put(851,448.84){\rule{1.686pt}{0.800pt}}
\multiput(851.00,448.34)(3.500,1.000){2}{\rule{0.843pt}{0.800pt}}
\put(858,449.84){\rule{1.445pt}{0.800pt}}
\multiput(858.00,449.34)(3.000,1.000){2}{\rule{0.723pt}{0.800pt}}
\put(864,450.84){\rule{1.445pt}{0.800pt}}
\multiput(864.00,450.34)(3.000,1.000){2}{\rule{0.723pt}{0.800pt}}
\put(870,451.84){\rule{1.686pt}{0.800pt}}
\multiput(870.00,451.34)(3.500,1.000){2}{\rule{0.843pt}{0.800pt}}
\put(877,452.84){\rule{1.445pt}{0.800pt}}
\multiput(877.00,452.34)(3.000,1.000){2}{\rule{0.723pt}{0.800pt}}
\put(883,453.84){\rule{1.445pt}{0.800pt}}
\multiput(883.00,453.34)(3.000,1.000){2}{\rule{0.723pt}{0.800pt}}
\put(838.0,449.0){\rule[-0.400pt]{1.686pt}{0.800pt}}
\put(896,454.84){\rule{1.445pt}{0.800pt}}
\multiput(896.00,454.34)(3.000,1.000){2}{\rule{0.723pt}{0.800pt}}
\put(902,455.84){\rule{1.686pt}{0.800pt}}
\multiput(902.00,455.34)(3.500,1.000){2}{\rule{0.843pt}{0.800pt}}
\put(909,456.84){\rule{1.445pt}{0.800pt}}
\multiput(909.00,456.34)(3.000,1.000){2}{\rule{0.723pt}{0.800pt}}
\put(915,457.84){\rule{1.445pt}{0.800pt}}
\multiput(915.00,457.34)(3.000,1.000){2}{\rule{0.723pt}{0.800pt}}
\put(889.0,456.0){\rule[-0.400pt]{1.686pt}{0.800pt}}
\put(928,458.84){\rule{1.445pt}{0.800pt}}
\multiput(928.00,458.34)(3.000,1.000){2}{\rule{0.723pt}{0.800pt}}
\put(934,459.84){\rule{1.686pt}{0.800pt}}
\multiput(934.00,459.34)(3.500,1.000){2}{\rule{0.843pt}{0.800pt}}
\put(941,460.84){\rule{1.445pt}{0.800pt}}
\multiput(941.00,460.34)(3.000,1.000){2}{\rule{0.723pt}{0.800pt}}
\put(921.0,460.0){\rule[-0.400pt]{1.686pt}{0.800pt}}
\put(953,461.84){\rule{1.686pt}{0.800pt}}
\multiput(953.00,461.34)(3.500,1.000){2}{\rule{0.843pt}{0.800pt}}
\put(960,462.84){\rule{1.445pt}{0.800pt}}
\multiput(960.00,462.34)(3.000,1.000){2}{\rule{0.723pt}{0.800pt}}
\put(966,463.84){\rule{1.686pt}{0.800pt}}
\multiput(966.00,463.34)(3.500,1.000){2}{\rule{0.843pt}{0.800pt}}
\put(947.0,463.0){\rule[-0.400pt]{1.445pt}{0.800pt}}
\put(979,464.84){\rule{1.445pt}{0.800pt}}
\multiput(979.00,464.34)(3.000,1.000){2}{\rule{0.723pt}{0.800pt}}
\put(985,465.84){\rule{1.686pt}{0.800pt}}
\multiput(985.00,465.34)(3.500,1.000){2}{\rule{0.843pt}{0.800pt}}
\put(973.0,466.0){\rule[-0.400pt]{1.445pt}{0.800pt}}
\put(998,466.84){\rule{1.445pt}{0.800pt}}
\multiput(998.00,466.34)(3.000,1.000){2}{\rule{0.723pt}{0.800pt}}
\put(1004,467.84){\rule{1.686pt}{0.800pt}}
\multiput(1004.00,467.34)(3.500,1.000){2}{\rule{0.843pt}{0.800pt}}
\put(992.0,468.0){\rule[-0.400pt]{1.445pt}{0.800pt}}
\put(1017,468.84){\rule{1.686pt}{0.800pt}}
\multiput(1017.00,468.34)(3.500,1.000){2}{\rule{0.843pt}{0.800pt}}
\put(1024,469.84){\rule{1.445pt}{0.800pt}}
\multiput(1024.00,469.34)(3.000,1.000){2}{\rule{0.723pt}{0.800pt}}
\put(1011.0,470.0){\rule[-0.400pt]{1.445pt}{0.800pt}}
\put(1036,470.84){\rule{1.686pt}{0.800pt}}
\multiput(1036.00,470.34)(3.500,1.000){2}{\rule{0.843pt}{0.800pt}}
\put(1043,471.84){\rule{1.445pt}{0.800pt}}
\multiput(1043.00,471.34)(3.000,1.000){2}{\rule{0.723pt}{0.800pt}}
\put(1030.0,472.0){\rule[-0.400pt]{1.445pt}{0.800pt}}
\put(1056,472.84){\rule{1.445pt}{0.800pt}}
\multiput(1056.00,472.34)(3.000,1.000){2}{\rule{0.723pt}{0.800pt}}
\put(1049.0,474.0){\rule[-0.400pt]{1.686pt}{0.800pt}}
\put(1068,473.84){\rule{1.686pt}{0.800pt}}
\multiput(1068.00,473.34)(3.500,1.000){2}{\rule{0.843pt}{0.800pt}}
\put(1075,474.84){\rule{1.445pt}{0.800pt}}
\multiput(1075.00,474.34)(3.000,1.000){2}{\rule{0.723pt}{0.800pt}}
\put(1062.0,475.0){\rule[-0.400pt]{1.445pt}{0.800pt}}
\put(1088,475.84){\rule{1.445pt}{0.800pt}}
\multiput(1088.00,475.34)(3.000,1.000){2}{\rule{0.723pt}{0.800pt}}
\put(1081.0,477.0){\rule[-0.400pt]{1.686pt}{0.800pt}}
\put(1100,476.84){\rule{1.686pt}{0.800pt}}
\multiput(1100.00,476.34)(3.500,1.000){2}{\rule{0.843pt}{0.800pt}}
\put(1094.0,478.0){\rule[-0.400pt]{1.445pt}{0.800pt}}
\put(1113,477.84){\rule{1.445pt}{0.800pt}}
\multiput(1113.00,477.34)(3.000,1.000){2}{\rule{0.723pt}{0.800pt}}
\put(1119,478.84){\rule{1.686pt}{0.800pt}}
\multiput(1119.00,478.34)(3.500,1.000){2}{\rule{0.843pt}{0.800pt}}
\put(1107.0,479.0){\rule[-0.400pt]{1.445pt}{0.800pt}}
\put(1132,479.84){\rule{1.686pt}{0.800pt}}
\multiput(1132.00,479.34)(3.500,1.000){2}{\rule{0.843pt}{0.800pt}}
\put(1126.0,481.0){\rule[-0.400pt]{1.445pt}{0.800pt}}
\put(1145,480.84){\rule{1.445pt}{0.800pt}}
\multiput(1145.00,480.34)(3.000,1.000){2}{\rule{0.723pt}{0.800pt}}
\put(1139.0,482.0){\rule[-0.400pt]{1.445pt}{0.800pt}}
\put(1158,481.84){\rule{1.445pt}{0.800pt}}
\multiput(1158.00,481.34)(3.000,1.000){2}{\rule{0.723pt}{0.800pt}}
\put(1151.0,483.0){\rule[-0.400pt]{1.686pt}{0.800pt}}
\put(1171,482.84){\rule{1.445pt}{0.800pt}}
\multiput(1171.00,482.34)(3.000,1.000){2}{\rule{0.723pt}{0.800pt}}
\put(1164.0,484.0){\rule[-0.400pt]{1.686pt}{0.800pt}}
\put(1183,483.84){\rule{1.686pt}{0.800pt}}
\multiput(1183.00,483.34)(3.500,1.000){2}{\rule{0.843pt}{0.800pt}}
\put(1177.0,485.0){\rule[-0.400pt]{1.445pt}{0.800pt}}
\put(1196,484.84){\rule{1.686pt}{0.800pt}}
\multiput(1196.00,484.34)(3.500,1.000){2}{\rule{0.843pt}{0.800pt}}
\put(1190.0,486.0){\rule[-0.400pt]{1.445pt}{0.800pt}}
\put(1209,485.84){\rule{1.445pt}{0.800pt}}
\multiput(1209.00,485.34)(3.000,1.000){2}{\rule{0.723pt}{0.800pt}}
\put(1203.0,487.0){\rule[-0.400pt]{1.445pt}{0.800pt}}
\put(1222,486.84){\rule{1.445pt}{0.800pt}}
\multiput(1222.00,486.34)(3.000,1.000){2}{\rule{0.723pt}{0.800pt}}
\put(1215.0,488.0){\rule[-0.400pt]{1.686pt}{0.800pt}}
\put(1235,487.84){\rule{1.445pt}{0.800pt}}
\multiput(1235.00,487.34)(3.000,1.000){2}{\rule{0.723pt}{0.800pt}}
\put(1228.0,489.0){\rule[-0.400pt]{1.686pt}{0.800pt}}
\put(1247,488.84){\rule{1.686pt}{0.800pt}}
\multiput(1247.00,488.34)(3.500,1.000){2}{\rule{0.843pt}{0.800pt}}
\put(1241.0,490.0){\rule[-0.400pt]{1.445pt}{0.800pt}}
\put(1260,489.84){\rule{1.445pt}{0.800pt}}
\multiput(1260.00,489.34)(3.000,1.000){2}{\rule{0.723pt}{0.800pt}}
\put(1254.0,491.0){\rule[-0.400pt]{1.445pt}{0.800pt}}
\put(1273,490.84){\rule{1.445pt}{0.800pt}}
\multiput(1273.00,490.34)(3.000,1.000){2}{\rule{0.723pt}{0.800pt}}
\put(1266.0,492.0){\rule[-0.400pt]{1.686pt}{0.800pt}}
\put(1292,491.84){\rule{1.445pt}{0.800pt}}
\multiput(1292.00,491.34)(3.000,1.000){2}{\rule{0.723pt}{0.800pt}}
\put(1279.0,493.0){\rule[-0.400pt]{3.132pt}{0.800pt}}
\put(1305,492.84){\rule{1.445pt}{0.800pt}}
\multiput(1305.00,492.34)(3.000,1.000){2}{\rule{0.723pt}{0.800pt}}
\put(1298.0,494.0){\rule[-0.400pt]{1.686pt}{0.800pt}}
\put(1318,493.84){\rule{1.445pt}{0.800pt}}
\multiput(1318.00,493.34)(3.000,1.000){2}{\rule{0.723pt}{0.800pt}}
\put(1311.0,495.0){\rule[-0.400pt]{1.686pt}{0.800pt}}
\put(1337,494.84){\rule{1.445pt}{0.800pt}}
\multiput(1337.00,494.34)(3.000,1.000){2}{\rule{0.723pt}{0.800pt}}
\put(1324.0,496.0){\rule[-0.400pt]{3.132pt}{0.800pt}}
\put(1350,495.84){\rule{1.445pt}{0.800pt}}
\multiput(1350.00,495.34)(3.000,1.000){2}{\rule{0.723pt}{0.800pt}}
\put(1343.0,497.0){\rule[-0.400pt]{1.686pt}{0.800pt}}
\put(1369,496.84){\rule{1.445pt}{0.800pt}}
\multiput(1369.00,496.34)(3.000,1.000){2}{\rule{0.723pt}{0.800pt}}
\put(1356.0,498.0){\rule[-0.400pt]{3.132pt}{0.800pt}}
\put(1381,497.84){\rule{1.686pt}{0.800pt}}
\multiput(1381.00,497.34)(3.500,1.000){2}{\rule{0.843pt}{0.800pt}}
\put(1375.0,499.0){\rule[-0.400pt]{1.445pt}{0.800pt}}
\put(1401,498.84){\rule{1.445pt}{0.800pt}}
\multiput(1401.00,498.34)(3.000,1.000){2}{\rule{0.723pt}{0.800pt}}
\put(1388.0,500.0){\rule[-0.400pt]{3.132pt}{0.800pt}}
\put(1420,499.84){\rule{1.445pt}{0.800pt}}
\multiput(1420.00,499.34)(3.000,1.000){2}{\rule{0.723pt}{0.800pt}}
\put(1407.0,501.0){\rule[-0.400pt]{3.132pt}{0.800pt}}
\put(1433,500.84){\rule{1.445pt}{0.800pt}}
\multiput(1433.00,500.34)(3.000,1.000){2}{\rule{0.723pt}{0.800pt}}
\put(1426.0,502.0){\rule[-0.400pt]{1.686pt}{0.800pt}}
\sbox{\plotpoint}{\rule[-0.500pt]{1.000pt}{1.000pt}}%
\put(1279,697){\makebox(0,0)[r]{$p=1$}}
\multiput(1299,697)(20.756,0.000){5}{\usebox{\plotpoint}}
\put(1399,697){\usebox{\plotpoint}}
\put(161,172){\usebox{\plotpoint}}
\put(161.00,172.00){\usebox{\plotpoint}}
\put(177.75,184.13){\usebox{\plotpoint}}
\put(193.09,198.09){\usebox{\plotpoint}}
\put(207.90,212.54){\usebox{\plotpoint}}
\put(221.42,228.22){\usebox{\plotpoint}}
\put(233.94,244.78){\usebox{\plotpoint}}
\put(245.71,261.85){\usebox{\plotpoint}}
\put(257.11,279.18){\usebox{\plotpoint}}
\put(267.99,296.85){\usebox{\plotpoint}}
\put(278.15,314.94){\usebox{\plotpoint}}
\put(288.83,332.73){\usebox{\plotpoint}}
\put(298.77,350.92){\usebox{\plotpoint}}
\put(308.59,369.18){\usebox{\plotpoint}}
\put(318.65,387.30){\usebox{\plotpoint}}
\put(328.59,405.49){\usebox{\plotpoint}}
\put(338.77,423.55){\usebox{\plotpoint}}
\put(349.41,441.36){\usebox{\plotpoint}}
\put(360.24,459.06){\usebox{\plotpoint}}
\put(371.59,476.42){\usebox{\plotpoint}}
\put(383.67,493.29){\usebox{\plotpoint}}
\put(395.66,510.22){\usebox{\plotpoint}}
\put(409.61,525.55){\usebox{\plotpoint}}
\put(424.26,540.26){\usebox{\plotpoint}}
\put(439.85,553.85){\usebox{\plotpoint}}
\put(456.53,566.02){\usebox{\plotpoint}}
\put(474.55,576.28){\usebox{\plotpoint}}
\put(493.31,585.13){\usebox{\plotpoint}}
\put(512.80,592.23){\usebox{\plotpoint}}
\put(532.86,597.29){\usebox{\plotpoint}}
\put(553.07,601.69){\usebox{\plotpoint}}
\put(573.46,605.00){\usebox{\plotpoint}}
\put(594.02,607.72){\usebox{\plotpoint}}
\put(614.67,609.00){\usebox{\plotpoint}}
\put(635.27,611.00){\usebox{\plotpoint}}
\put(655.98,611.50){\usebox{\plotpoint}}
\put(676.70,612.00){\usebox{\plotpoint}}
\put(697.37,613.00){\usebox{\plotpoint}}
\put(718.13,613.00){\usebox{\plotpoint}}
\put(738.80,614.00){\usebox{\plotpoint}}
\put(759.55,614.00){\usebox{\plotpoint}}
\put(780.31,614.00){\usebox{\plotpoint}}
\put(801.07,614.00){\usebox{\plotpoint}}
\put(821.82,614.00){\usebox{\plotpoint}}
\put(842.58,614.00){\usebox{\plotpoint}}
\put(863.33,614.00){\usebox{\plotpoint}}
\put(884.09,614.00){\usebox{\plotpoint}}
\put(904.84,614.00){\usebox{\plotpoint}}
\put(925.60,614.00){\usebox{\plotpoint}}
\put(946.35,614.00){\usebox{\plotpoint}}
\put(967.11,614.00){\usebox{\plotpoint}}
\put(987.86,614.00){\usebox{\plotpoint}}
\put(1008.62,614.00){\usebox{\plotpoint}}
\put(1029.38,614.00){\usebox{\plotpoint}}
\put(1050.13,614.00){\usebox{\plotpoint}}
\put(1070.89,614.00){\usebox{\plotpoint}}
\put(1091.64,614.00){\usebox{\plotpoint}}
\put(1112.40,614.00){\usebox{\plotpoint}}
\put(1133.15,614.00){\usebox{\plotpoint}}
\put(1153.91,614.00){\usebox{\plotpoint}}
\put(1174.66,614.00){\usebox{\plotpoint}}
\put(1195.42,614.00){\usebox{\plotpoint}}
\put(1216.17,614.00){\usebox{\plotpoint}}
\put(1236.93,614.00){\usebox{\plotpoint}}
\put(1257.69,614.00){\usebox{\plotpoint}}
\put(1278.44,614.00){\usebox{\plotpoint}}
\put(1299.20,614.00){\usebox{\plotpoint}}
\put(1319.95,614.00){\usebox{\plotpoint}}
\put(1340.71,614.00){\usebox{\plotpoint}}
\put(1361.46,614.00){\usebox{\plotpoint}}
\put(1382.22,614.00){\usebox{\plotpoint}}
\put(1402.97,614.00){\usebox{\plotpoint}}
\put(1423.73,614.00){\usebox{\plotpoint}}
\put(1439,614){\usebox{\plotpoint}}
\end{picture}

\end{center}
\caption{Time evolutions of fractional logistic map}
\end{figure}
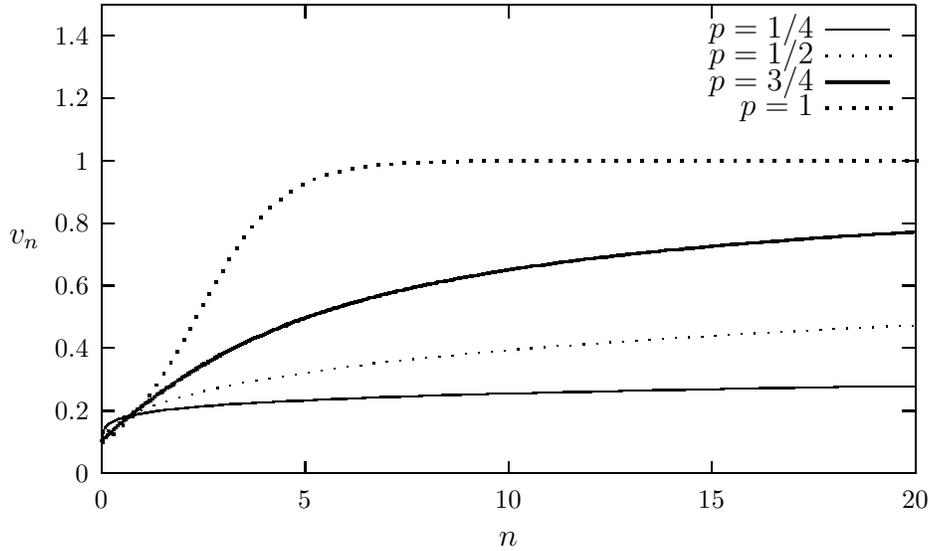

Considering the fact that the Mittag-Leffler function has an 
asymptotic behavior~\cite{Sansonne},
\begin{align}
E_a(\lambda t^a) = -\sum_{n=1}^{N-1}\frac{\lambda^{-n}t^{-an}}
{\Gamma(1-an)}+O(t^{-aN}), \quad t\rightarrow\infty, \lambda<0,
\end{align}
we can observe that $u_n$ converges to $1$ at the order 
of $O(1/n^p)$ if $0<p<1$.
The following table illustrates a numerical result in which we 
apply the $\rho$-algorithm~\cite{Brezinski}
\begin{align*}
\rho_{k+1}^n &= \rho_{k-1}^{n+1}+\frac{(k+n)^p-n^p}{
\rho_k^{n+1}-\rho_k^n}\\
& \rho_{-1}^n = 0, \ \rho_0^n = u_n
\end{align*}
to the sequence $\{u_n\}$ in the case $p=1/4$. This table shows
that $u_n$ converges to the value near $1.0$ at the order 
$O(1/n^{1/4})$ as $n$ tends to $+\infty$.

\begin{table}[htbp]
\caption{The $\rho$-algorithm applied to the sequence $\{u_n\}$ 
in the case $p=1/4$}
\begin{center}
\begin{tabular}{|l||l|l|l|l|l|l|}\hline
& $\rho_1^n$ & $\rho_3^n$ & $\rho_5^n$ & $\rho_7^n$ & $\cdots$ &
$\rho_{21}^n$ \\ \hline
1 & 0.1    & 0.19745 & 0.84609 & 0.84288  && 1.00145\\ \hline
2 & 0.14791 & 0.23921 & 0.86006 & 1.30904 && 0.99915\\ \hline
3 & 0.16019 & 0.27802 & 0.99118 & 1.21129 && 1.00131\\ \hline
4 & 0.16764 & 0.31406 & 1.07379&  1.17824 && 1.00139\\ \hline
5 & 0.17313 & 0.34778 & 1.11866 & 1.16134 && 1.00112\\ \hline
6 & 0.17752 & 0.37941 & 1.14064 & 1.15284 && 1.00120\\ \hline
7 & 0.18121 & 0.40908 & 1.14968  & 1.15045 && 1.00120 \\ \hline
8 & 0.18442 & 0.43692 & 1.15161  & 1.15375 && 1.00121 \\ \hline
9 & 0.18726 & 0.46304 & 1.14970  & 1.16376 && 1.00122 \\ \hline
\vdots & \vdots & \vdots & \vdots & \vdots && \vdots \\ \hline
\end{tabular}
\end{center}
\end{table}
\section*{Acknowledgement}
The author would like to express his sincere thanks to Professor 
Yoshinori Kametaka at Osaka University for providing me with 
information of Mittag-Leffler function. He also thanks Professor
Junkichi Satsuma at University of Tokyo for many valuable suggestions.

\end{document}